\documentclass[12pt]{iopart}
\usepackage{iopams}
\usepackage{setstack}
\usepackage{psfig}

\newcommand{\half}{\mbox{$\textstyle \frac{1}{2}$}}

\newcommand{\ri}{{\rm i}}
\newcommand{\rd}{{\rm d}}
\begin{document}

\title[Solvable model of quantum microcanonical states]{Solvable model of
quantum microcanonical states}

\author[Bender, Brody, and Hook]{Carl~M~Bender$^{*}$, Dorje~C~Brody$^{\dagger}$,
and Daniel~W~Hook$^{\dagger}$}

\address{${}^{*}$Department of Physics, Washington University, St. Louis MO
63130, USA}

\address{${}^{\dagger}$Blackett Laboratory, Imperial College, London SW7 2BZ,
UK}

\begin{abstract}
This letter examines the consequences of a recently proposed
modification of the postulate of equal {\it a priori} probability
in quantum statistical mechanics. This modification, called the
{\it quantum microcanonical postulate} (QMP), asserts that for a
system in microcanonical equilibrium all pure quantum states
having the same energy expectation value are realised with equal
probability. A simple model of a quantum system that obeys the QMP
and that has a nondegenerate spectrum with equally spaced energy
eigenvalues is studied. This model admits a closed-form expression
for the density of states in terms of the energy eigenvalues. It
is shown that in the limit as the number of energy levels
approaches infinity, the expression for the density of states
converges to a $\delta$ function centred at the intermediate value
$(E_{\rm max}+E_{\rm min})/ 2$ of the energy. Determining this
limit requires an elaborate asymptotic study of an infinite sum
whose terms alternate in sign.
\end{abstract}

\submitto{\JPA}

\section{Introduction}\label{sec:1}
This letter investigates a generalization of the usual definition
of a quantum system in microcanonical equilibrium. If the
Hamiltonian $H$ that describes a system has a nondegenerate
spectrum, then according to the standard definition of quantum
microcanonical equilibrium the system must be in one of the
eigenstates of $H$. This requirement is known as the {\it
postulate of equal a priori probabilities}~\cite{huang}. We
emphasize that according to the definition of microcanonical
equilibrium in Ref.~\cite{huang} the state of such a system cannot
be a linear combination of eigenstates of $H$. However, if $H$ has
a degenerate spectrum, then the density matrix that describes a
system of energy $E$ in microcanonical equilibrium contains all
states $|E,k\rangle$ of the degenerate energy $E$ with equal
weight: $\frac{1}{n}\sum_{k=1}^n|E,k\rangle \langle E,k|$, where
$n$ is the number of states having energy $E$.

Because the standard definition of quantum microcanonical
equilibrium only allows the system to have energies that are
eigenvalues of $H$, an alternative, less restrictive definition
has recently been introduced \cite{BHH}. By this latter
definition, called the {\it quantum microcanonical postulate}
(QMP), a state of a system in microcanonical equilibrium can have
an energy that is {\it not} an eigenvalue of $H$. The discussion
in Ref.~\cite{BHH} of quantum systems obeying the QMP is
qualitative. Here, we give a quantitative analysis of a quantum
system described by a Hamiltonian having a nondegenerate, equally
spaced spectrum. Assuming that the system is in microcanonical
equilibrium and that it obeys the QMP, we study the behaviour of
the density of states $\mu(E)$ as the number of energy levels
becomes large.

This letter is organised as follows: In Sec.~\ref{sec:2} we review
the representation for the density of states $\mu(E)$ in terms of
the energy eigenvalues as outlined in Ref.~\cite{BHH}. We then
define the model investigated in this letter in which the energy
spectrum is taken to be nondegenerate and to grow linearly:
$E_k\propto k$. In the next two sections we investigate the
behaviour of $\mu(E)$ as the number of states of $H$ becomes
infinite. In Sec.~\ref{sec:3} we show that $\mu(E)$ integrates to
unity. Section~\ref{sec:4} presents an asymptotic study of
$\mu(E)$ as the number of states becomes infinite. On the basis of
the analysis given in Secs.~\ref{sec:3} and \ref{sec:4}, we
conclude that $\mu(E)$ approaches $\delta[E-(E_{{\rm max}}+
E_{{\rm min}})/2]$.

\section{Definition of the model}
\label{sec:2} Let us review briefly the general mathematical
framework proposed in Ref.~\cite{BHH} for describing the density
matrix of a mixed state of a quantum system in microcanonical
equilibrium. Consider a quantum system defined on an
$(n+1)$-dimensional Hilbert space ${\mathcal H}$. Let $Z^\alpha$
$(\alpha=0,1,2, \ldots,n)$ be a typical element of ${\mathcal H}$
and let $H^\alpha_\beta$ denote the Hamiltonian with eigenvalues
$E_i$ $(i=0,1,2,\ldots,n)$. Then, the expectation value of the
Hamiltonian in the state $Z^\alpha$ is $\langle H
\rangle=\bar{Z}_\alpha H^\alpha_\beta Z^\beta/\bar{Z}_\gamma
Z^\gamma$. Assume that in microcanonical equilibrium all states
$Z^\alpha$ satisfying the condition $\langle H\rangle=E$ are
realised with equal probability. Then, the corresponding
unnormalised density of states $\Omega(E)$ is
\begin{eqnarray}
\Omega(E)=\frac{1}{\pi}\int_{\mathcal
H}{\rd}^{n+1}\bar{Z}\,{\rd}^{n+1}Z\,\delta (\bar{Z}_\alpha
Z^\alpha-1)\;\delta\left(\frac{\bar{Z}_\alpha H^\alpha_\beta
Z^\beta}{\bar{Z}_\gamma Z^\gamma}-E\right). \label{eq:2}
\end{eqnarray}

The constraint $\delta(\bar{Z}_\alpha Z^\alpha-1)$ in (\ref{eq:2})
arises because one is only interested in the unit normalised
states, and the factor of $\pi$ reflects the additional redundant
overall phase of the state. It is convenient to use the standard
integral representation $\delta(x)=\frac{1}{2\pi}
\int_{-\infty}^{\infty}{\rd}\lambda\,e^{-\ri\lambda x}$ for each
of the $\delta$ functions appearing in (\ref{eq:2}). The
Hilbert-space integration then becomes Gaussian in the $Z$
variables leaving the expression
\begin{eqnarray}
\Omega(E)=\frac{1}{\pi}(-\ri\pi)^{n+1}\int_{-\infty}^{\infty}\frac{{\rd}
\nu}{2\pi}\int_{-\infty}^{\infty}\frac{{\rd}\lambda}{2\pi}\,e^{\ri
(\lambda+\nu E)}\prod_{l=0}^{n}\frac{1}{\lambda+\nu E_l}.
\label{eq:4}
\end{eqnarray}

Assuming that the energy spectrum is nondegenerate, one can
perform the $\lambda$ integration to obtain
\begin{eqnarray}
\Omega(E)=\pi^{n}\sum_{k=0}^{n}\int_{-\infty}^{\infty}\frac{{\rd}\nu}{2\pi}
\frac{e^{-\ri\nu(E_k-E)}}{(\ri\nu)^n}\prod_{l=0,\neq k}^{n}
\frac{1}{E_l-E_k}. \label{eq:5}
\end{eqnarray}
The remaining $\nu$ integration can now be performed explicitly to
give
\begin{eqnarray}
\Omega(E)=(-\pi)^n\sum_{k=0}^n\delta^{(-n)}(E_k-E)\prod_{l=0,\neq
k}^{n}\frac{1} {E_l-E_k}, \label{eq:6}
\end{eqnarray}
where $\delta^{(-n)}(x)$ denotes the $n$th integral of the
$\delta$ function:
\begin{eqnarray}
\delta^{(-n)}(x)= \left\{ \begin{array}{ll} 0 & (x<0), \\
\frac{1}{(n-1)!}\, x^{n-1} & (x\geq 0) .\end{array}\right.
\label{eq:7}
\end{eqnarray}

The density of states $\Omega(E)$ as defined in (\ref{eq:2}) is
normalised by dividing it by the volume of the subspace of
${\mathcal H}$ spanned by states having unit length: ${\bar
Z}_\alpha Z^\alpha=1$. This gives the normalised microcanonical
state density function $\mu(E)$. The volume is given by $\pi^n/
n!$ (see, for example, Ref.~\cite{gibbons}). Thus,
$\mu(E)=n!\,\pi^{-n}\,\Omega( E)$ gives the density of states that
satisfies the normalisation condition $\int_{-\infty}^{\infty}\rd
E\,\mu(E)=1$.

In this letter we propose a particular QMP model in which the
energy spectrum rises linearly and is given by $E_k=k$. Our
objective is to study the behaviour of $\mu(E)$ as the number of
energy levels becomes infinite. With this linear choice of
spectrum the normalised density of states becomes
\begin{eqnarray}
\mu(E)=(-1)^n
n\sum_{k>[E]}^n\frac{(-1)^k\left(k-E\right)^{n-1}}{k!(n-k)!},
\label{eq:9}
\end{eqnarray}
where the notation $[E]$ indicates the largest integer less than
or equal to $E$.

It is now convenient to rescale the energy spectrum so that the
range of the energy lies in the interval $[0,1]$ for each $n$.
Upon rescaling, (\ref{eq:9}) transforms to
\begin{eqnarray}
\mu(E)=(-1)^{n+1}n^2\sum_{k=0}^{[nE]}\frac{(-1)^k(k-nE)^{n-1}}{k!(n-k)!},
\label{eq:10}
\end{eqnarray}
where $E\in[0,1]$ for all $n$. To derive this result we have used
the fact that the sum in (\ref{eq:9}) vanishes when the summation
range is taken from $k=0$ to $k=n$.

\begin{figure}[th]\vspace{2.6in}
\includegraphics{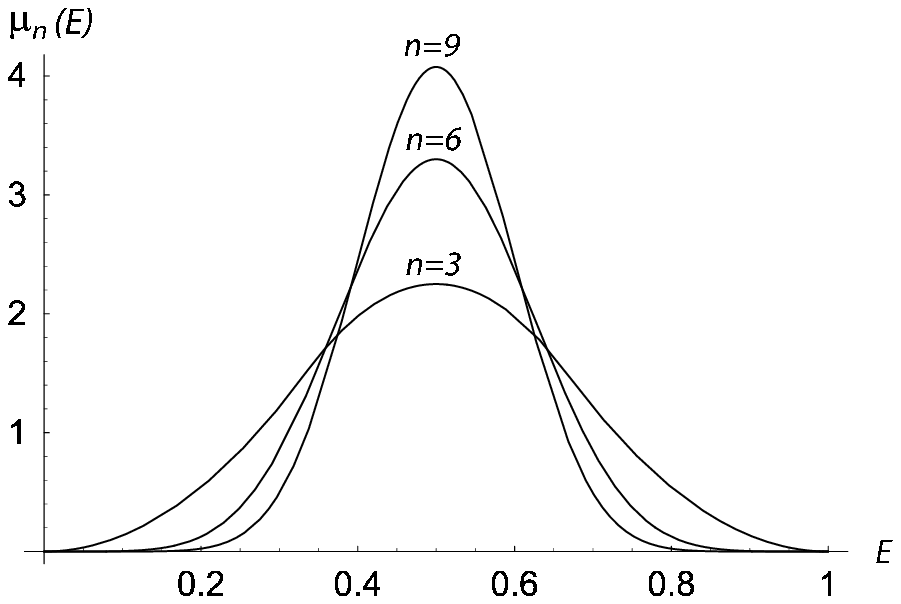} \caption{The density of states $\mu_n(E)$
associated with a quantum system having a linear energy spectrum
$E_k=k$, where the range of the energy is suitably rescaled so
that $E$ lies in the range $[0,1]$ for all $n$. Plots of
$\mu_n(E)$ are given for 4-, 7-, and 10-state ($n=3$, $6$, and
$9$) systems. Observe that as the number of energy levels
increases, the distribution becomes more peaked at the centre
$E=1/2$, suggesting that as the number of energy levels approaches
infinity, the distribution approaches $\delta(E-1/2)$. The
analysis in Secs.~\ref{sec:3} and \ref{sec:4} verifies that this
is indeed the case. \label{fig:1}}
\end{figure}

In Fig.~\ref{fig:1} we plot the density of states $\mu(E)$ in
(\ref{eq:10}) for $n=3,~6$, and $9$. This graph suggests that
$\mu(E)$ converges to a $\delta$ function centred at $E=1/2$ as
$n$, the number of energy levels, increases. We show analytically
that the density of states $\mu(E)$ associated with a quantum
system having the spectrum $E_k\propto k$ does indeed approach
$\delta(E-1/2)$ in the limit $n\to\infty$. Our analysis is of
interest because it involves an asymptotic study of an infinite
sum whose terms alternate in sign. To overcome the difficulties
associated with this alternating series, we convert the series to
a double contour integration whose asymptotic behaviour is
obtained using the method of steepest descent. This work also
provides a new limit identity for the Dirac $\delta$ function.

\section{Analysis of the model}
\label{sec:3} To verify that $\mu(E)$ approaches $\delta(E-1/2)$
as the number of states $n$ approaches infinity, we must establish
two properties of $\mu(E)$. First, we must show that
$\int_{-\infty}^\infty\rd E\,\mu(E)=1$. Second, we must show that
the limiting value of $\mu(E)$ is zero except at $E=1/2$ where it
tends to infinity. In this section we show that the normalisation
condition satisfied by $\mu(E)$ in (\ref{eq:10}) is valid. Let us
define $I$ by $I=\int_0^1\rd E\,\mu(E )$. The summation in
(\ref{eq:10}) must be evaluated piecewise because of the
dependence of the summation range on $E$. Thus, it is convenient
to decompose the integration range of $I$ into $n$ intervals and
to write
\begin{eqnarray}
I=(-1)^{n+1}n^2\sum_{j=1}^{n}\int_{(j-1)/n}^{j/n}\rd
E\sum_{k=0}^{[nE]}\frac{ (-1)^k(k-nE)^{n-1}}{k!(n-k)!}.
\label{eq:x5}
\end{eqnarray}

To perform the integration over $E$ we rewrite the summation in
the integrand so that it is independent of $E$. Given that $j$
ranges from $1$ to $n$ and that $k$ ranges from $0$ to $n-1$ with
$k\leq j-1$, we have
\begin{eqnarray}
I &=& (-1)^{n+1} n^2
\sum_{j=1}^n\sum_{k=0}^{j-1}\int_{(j-1)/n}^{j/n}\;{\rd}E
\frac{(-1)^k(k-nE)^{n-1}}{k!(n-k)!}\nonumber\\
&=& (-1)^{n+1} n^2 \sum_{j=1}^n \sum_{k=0}^{j-1}
\frac{(-1)^k}{k!(n-k)!} \left.
\frac{(k-nE)^n}{n(-n)}\right|_{E=(j-1)/n}^{j/n}.
\end{eqnarray}
We now interchange the order of summation according to
$\sum_{j=1}^n\sum_{k=0}^{ j-1}=\sum_{k=0}^n\sum_{j=k+1}^n$:
\begin{eqnarray}
\label{eq:x6}
I=\sum_{k=0}^{n-1}\frac{(-1)^k}{k!(n-k)!}\sum_{j=k+1}^{n}\left[(j-k)^n-(j-k-1)^n
\right].
\end{eqnarray}
Performing the sum over $j$, we obtain
\begin{eqnarray}
\label{eq:x9}
I=\sum_{k=0}^{n}\frac{(-1)^k}{k!(n-k)!}(n-k)^n=\frac{(-1)^n}{n!}\sum_{k=0}^n(-1)
^k\left({n\atop k}\right)(k-n)^n.
\end{eqnarray}

Observe that the summation in (\ref{eq:x9}) is the $n$th discrete
difference of $k^n$. Recall that for the polynomial $f(k)=k^n+{\rm
lower~powers}$, the first discrete difference is
$\mathcal{D}f(k)=f(k)-f(k-1)=nk^{n-1}+{\rm lower~ powers}$. The
second discrete difference is $\mathcal{D}^2f(k)=f(k)-2f(k-1)+f(k-
2)=n(n-1)k^{n-2}+{\rm lower~powers}$, and so on. The $n$th
discrete difference is especially simple because there are no
remaining lower powers: $\mathcal{D}^n f(k)=n!$. This observation
allows us to evaluate the sum in (\ref{eq:x9}) :
\begin{eqnarray}
\label{eq:x12} \sum_{k=0}^n \left({n \atop k}\right)(-1)^k
k^n=(-1)^n n!.
\end{eqnarray}
We have thus verified the normalization condition $I=1$.

\section{Asymptotic behaviour of (\ref{eq:10}) for large $n$}
\label{sec:4} We now examine the behaviour of $\mu(E)$ in the
limit as $n\to\infty$. We have already shown in Sec.~\ref{sec:3}
that the integral of $\mu(E)$ is unity. To show that $\mu(E)$
approaches a delta function as $n\to\infty$ we must establish that
$\mu(E)$ becomes singular at the central value $E=\half(E_{{\rm
max}}+ E_{{\rm min}})$ and that it vanishes at all other points in
this limit. Note that by the scaling used in (\ref{eq:10}) the
central value is at $E=1/2$.

The representation of $\mu(E)$ in (\ref{eq:10}) for finite values
of $n$ is symmetric about the point $E=1/2$. To verify this
symmetry we make the transformation $E\to1-E$ and replace the
summation variable $k$ by $n-k$. Thus, we need only study the
behaviour of $\mu(E)$ for $E=1/\alpha$, where $\alpha\geq 2$.
Without loss of generality, we set $n=\alpha J$, where $J$ is a
large integer, and let $\omega_J(\alpha)$ be the value of $\mu(E)$
at $E=1/\alpha$:
\begin{eqnarray}
\label{eq:x14}
\omega_J(\alpha)=\alpha^2J^2\sum_{k=0}^J\frac{(-1)^k(J-k)^{\alpha
J-1}}{k!( \alpha J-k)!}.
\end{eqnarray}

It is straightforward to find the behaviour of $\omega_J(\alpha)$
for large $J$ when $\alpha>e$. Using Stirling's formula for the
asymptotic behaviour of the factorial function, we observe that
each term in the sum in (\ref{eq:x14}) is exponentially small;
that is, it has the form $e^{-AJ}$ $(J\to\infty)$, where $A$ is a
positive constant. The number of terms in the sum grows linearly
with $J$. Thus, the sum vanishes as $J\to\infty$.

However, when $2\leq\alpha\leq e$, the terms in the sum
(\ref{eq:x14}) are exponentially large. In this case, the factor
of $(-1)^k$ in the summand gives rise to a deep global
cancellation among all the terms in the sum. When $\alpha> 2$,
this cancellation causes $\omega_J(\alpha)$ to vanish
exponentially for large $J$. The case $\alpha=2$ is special
because the sum does not vanish exponentially. We have performed
the sum on the right side of (\ref{eq:x14}) numerically for large
values of $J$ when $\alpha=2$ using Richardson
extrapolation~\cite{bender}. We find that
\begin{eqnarray}
\omega_J(2)\sim(1.9544100476\ldots)\sqrt{J}\quad(J\to\infty).
\label{eq:x14a}
\end{eqnarray}
Establishing these asymptotic results analytically is difficult.
Laplace's method for sums cannot be used to evaluate
$\omega_J(\alpha)$ because Laplace's method involves local
analysis and this method is inadequate when terms in the sum
alternate in sign.

To overcome this difficulty we convert the sum in (\ref{eq:x14})
to a double complex contour integral. We begin by substituting
$k=J-p$:
\begin{eqnarray}
\omega_J(\alpha)=\alpha^2J^2\sum_{p=1}^J\frac{(-1)^{J+p}p^{\alpha
J-1}}{\Gamma(J -p+1)\Gamma[(\alpha-1)J+p+1]}. \label{eq:x16}
\end{eqnarray}
We then use the identity
$\frac{1}{\Gamma(z)}=\frac{1}{2\pi\ri}\oint_C{\rd}t\, e^{t}t^{-z}$
to represent the $\Gamma$ functions in (\ref{eq:x16}). The contour
$C$ is infinite and encloses the negative real-$t$ axis; $C$ can
be taken to be a circle around the origin when $\alpha$ is an
integer. Rescaling the integration variable $t$ gives
\begin{eqnarray}
\omega_J(\alpha)=(-1)^J\frac{\alpha^2J^2}{(2\pi\ri)^2}\oint_C\oint_{C'}\rd
r\, \rd
s\,r^{-(\alpha-1)J-1}s^{-J-1}\sum_{p=1}^J\frac{1}{p}\left(-\frac{s}{r}e^{r+s
}\right)^p, \label{eq:x19}
\end{eqnarray}
where we have interchanged orders of integration and summation.

Integrating (\ref{eq:x19}) by parts with respect to $r$, we obtain
\begin{eqnarray}
\label{eq:x20}
\omega_J(\alpha)=(-1)^J\frac{\alpha^2J}{(2\pi\ri)^2}\oint_C\oint_{C'}\frac{\rd
r}{r}\frac{\rd
s}{s}r^{-(\alpha-1)J}s^{-J}\frac{r-1}{\alpha-1}\sum_{p=1}^J
\left(-\frac{s}{r}e^{r+s}\right)^p.
\end{eqnarray}
We also integrate (\ref{eq:x19}) by parts with respect to $s$:
\begin{eqnarray}
\label{eq:x21}
\omega_J(\alpha)=(-1)^J\frac{\alpha^2J}{(2\pi\ri)^2}\oint_C\oint_{C'}\frac{\rd
r}{r}\frac{\rd
s}{s}r^{-(\alpha-1)J}s^{-J}(s+1)\sum_{p=1}^J\left(-\frac{s}{r}
e^{r+s}\right)^p.
\end{eqnarray}
We then evaluate the finite geometric sums in (\ref{eq:x20}) and
(\ref{eq:x21}) using the identity $\sum_{p=1}^J
a^p=(a^{J+1}-a)/(a-1)$, where $a=-e^re^s s/r$. The representation
for $\omega_J(\alpha)$ simplifies when we combine the right side
of (\ref{eq:x20}) multiplied by $(\alpha-1)/\alpha$ and the right
side of (\ref{eq:x21}) multiplied by $1/\alpha$, and then replace
$s$ by $-s$:
\begin{eqnarray}
\omega_J(\alpha)=\frac{\alpha
J}{(2\pi\ri)^2}\oint_C\oint_{C'}{\rd}r\,{\rd}s\; (s-r)r^{-\alpha
J-1}\,e^r\frac{e^{J(r-s)}-r^Js^{-J}}{se^r-re^s}. \label{eq:28}
\end{eqnarray}

The term proportional to $e^{J(r-s)}$ in the integrand of
(\ref{eq:28}) is analytic in $s$ along the real-$s$ axis for
$s\leq0$. Hence, by shrinking the contour to a small circle about
the origin in the complex-$s$ plane, we find that the integrand
does not contribute to the asymptotic behaviour of (\ref{eq:28})
for large $J$. We have thus reduced the expression for $\omega_J(
\alpha)$ to
\begin{eqnarray}
\omega_J(\alpha)=\frac{\alpha
J}{(2\pi\ri)^2}\oint_C{\rd}r\,r^{-(\alpha-1)J-1}
\oint_{C'}{\rd}s\,s^{-J}\frac{(r-s)e^r}{se^r-re^s}. \label{eq:29}
\end{eqnarray}
Also, because the integral $\oint\rd s\,s^{-J}$ vanishes for
integer $J>1$, we may simplify (\ref{eq:29}) further by adding
$s^{-J}$ to the integrand of the $s$ integral:
\begin{eqnarray}
\omega_J(\alpha)=\frac{\alpha
J}{(2\pi\ri)^2}\oint_C{\rd}r\,r^{-(\alpha-1)J}
\oint_{C'}{\rd}s\,s^{-J}\,\frac{e^r-e^s}{se^r-re^s}. \label{eq:30}
\end{eqnarray}

Note that the integrand of (\ref{eq:30}) is singular if
\begin{eqnarray}
se^r-re^s=0, \label{eq:31}
\end{eqnarray}
as long as $e^r-e^s$ does not vanish. Clearly, (\ref{eq:31}) is
satisfied when $r=s$, but the numerator of the integrand in
(\ref{eq:30}) also vanishes when $r=s$. Thus, it may appear at
first that there is no singularity at $r=s$. However, for the
special point $r=s=1$ the denominator has a higher-order zero than
the numerator and thus the integrand is singular there. To find
the asymptotic behaviour of (\ref{eq:30}) we must perform a
steepest-descent analysis. However, if we look for a saddle-point
of the double integral, we find that it is located near the
singular point $r=1$ and $s=1$, which complicates the asymptotic
analysis enormously. Instead, we will evaluate the $s$ integral in
closed but implicit form and evaluate the remaining single
integral in $r$ asymptotically.

It is remarkable that the transcendental equation (\ref{eq:31})
has other solutions for which $r\neq s$. These solutions cannot be
expressed in closed form. However, we have discovered an explicit
parametric solution to (\ref{eq:31}) for which $r\neq s$:
\begin{eqnarray}
r=\lambda\,e^{-\lambda}/\sinh\lambda\quad{\rm and}\quad
s=\lambda\,e^\lambda/\sinh\lambda, \label{eq:32}
\end{eqnarray}
where $\lambda$ is any complex number \cite{rrr}.

To evaluate the $s$ integral in closed form we must take the
asymmetric solutions (\ref{eq:32}) into account. We treat the $C'$
contour as a circle about the origin in the complex-$s$ plane, but
rather than considering the singularities inside this circle, we
include instead the contributions of the singularities {\it
outside} this circle because the integrand vanishes at $|s|=
\infty$ in all directions. We now solve (\ref{eq:31}) for $s$ as a
function of $r$ and denote the solution as $s=S(r)$. We then use
residue calculus to evaluate the integral (\ref{eq:30}) at the
simple pole located at $s=S(r)$. The result is
\begin{eqnarray}
\omega_J(\alpha)=-\frac{\alpha
J}{2\pi\ri}\oint_C{\rd}r\,r^{-(\alpha-1)J-1}
[S(r)]^{-J}\frac{r-S(r)}{1-S(r)}, \label{eq:x25}
\end{eqnarray}
where we have simplified the integrand by using the algebraic
relation in (\ref{eq:31}).

To prepare for the asymptotic evaluation of the integral in
(\ref{eq:x25}) we rewrite it in standard Laplace form in terms of
the parametric variable $\lambda$ in (\ref{eq:32}):
\begin{eqnarray}
\omega_J(\alpha)=\oint{\rd}\lambda\,g(\lambda)e^{-Jf(\lambda)},
\end{eqnarray}
where $f(\lambda)=\log\left(\lambda
e^\lambda/\sinh\lambda\right)+(\alpha-1) \log\left(\lambda
e^{-\lambda}/\sinh\lambda\right)$ and $g(\lambda)=\alpha J
\sinh\lambda(1-\lambda-\lambda/\tanh\lambda)/[\ri\pi(\sinh\lambda-\lambda
e^\lambda)]$.

Following standard steepest-descent techniques~\cite{bender}, we
identify the saddle point as the solution $\lambda_0$ to
$f'(\lambda)=0$, where $f'(\lambda)=
2-\alpha+\alpha(1/\lambda-1/\tanh\lambda)$. It is easy to verify
that $f( \lambda_0)>0$ when $\alpha>2$. This implies that
$\omega_J(\alpha)$ vanishes exponentially rapidly like
$\omega_J(\alpha)\sim e^{-f(\lambda_0)J}$ as $J\to \infty$ for
$\alpha>2$. However, when $\alpha=2$, $\lambda_0=0$ and
$f(\lambda_0 )=0$. In this case $\omega_J(\alpha)$ behaves
algebraically for large $J$. To find this behaviour we calculate
$f''(\lambda_0)=-2/3$. Also, $g(\lambda_0)=-2 \ri J/\pi$. Thus,
the leading steepest-descent calculation shows that $\omega_J( 2)$
diverges as $J\to\infty$:
\begin{eqnarray}
\omega_J(2)\sim-\frac{2\ri
J}{\pi}\int{\rd}\lambda\,e^{2J\lambda^2/3}
\sim\frac{2\sqrt{3}}{\sqrt{\pi}}\sqrt{J}\quad(J\to\infty).
\end{eqnarray}
This reproduces the result of the Richardson extrapolation in
(\ref{eq:x14a}) and we identify
$\frac{2\sqrt{3}}{\sqrt{\pi}}=1.9544100476\ldots$.

In summary, we have shown that as the number of energy levels
increases, the normalised density of states $\mu(E)$ approaches
zero when $E\neq1/2$ and diverges when $E=1/2$. From this result
and the normalisation condition satisfied by $\mu(E)$, we conclude
that in this limit $\mu(E)\to\delta(E-1/2)$. Thus, according to
the postulate that every quantum state associated with a given
energy $E$ must be realised with equal probability in
microcanonical equilibrium, the density of states associated with
a system having a nondegenerate linear energy spectrum approaches
a delta function in the thermodynamic limit. It follows that in
this limit the energy of the system can assume only one value.
Whether an analogous result holds for an interacting system having
a degenerate spectrum is an interesting open question.

\vspace{0.5cm}
\begin{footnotesize}
\noindent CMB is supported by the US Department of Energy and DCB
is supported The Royal Society.
\end{footnotesize}

\end{document}